\documentclass[12pt,a4paper]{article}
\usepackage[USenglish]{babel}
\usepackage[T1]{fontenc}
\usepackage[ansinew]{inputenc}
\usepackage{lmodern}
\usepackage{multirow}
\usepackage[linktocpage,
   bookmarks=true,
   bookmarksnumbered=false,
   bookmarksopen=true,
   colorlinks=true,
   linkcolor=blue,
	 filecolor=green,
   urlcolor=red,
   citecolor=blue]{hyperref}
\usepackage{graphicx}
\usepackage{float}
\usepackage{amsmath}
\DeclareMathOperator{\sech}{sech}
\usepackage{amsfonts}
\usepackage{microtype}
\usepackage{setspace}
\usepackage{cleveref}
\crefname{equation}{equation}{equations}
\crefname{section}{Section}{Sections}
\crefname{chapter}{Chapter}{Chapters}
\crefname{table}{Table}{Tables}
\crefname{figure}{Figure}{Figures}
\usepackage{booktabs}

\usepackage[top=1in, left=1.2in, bottom=1in, right=1in]{geometry}
\usepackage[round]{natbib}
\usepackage{indentfirst}
\usepackage{setspace}
\title{\bf Paired Comparisons Modeling using \emph{t}--Distribution with Bayesian Analysis}
\begin{document}
\date{}
\definecolor{blue}{rgb}{0, 0, 0.8}
\floatstyle{boxed}
\restylefloat{figure}
\maketitle

\author \begin{center}
Maqsood Ali${}^a {}^{*}$, Muhammad Aslam${}^b$
				\end{center}
				\begin{center}
\footnotesize{\textsl{{${}^a$ Ludwig-Maximilians-Universit\"at, Munich, Germany.\\
${}^b$ Riphah International University, Islamabad, Pakistan.\\
${}^*$ E-mail: maqsoodfsd@outlook.com.}}}
				\end{center}

\section*{Abstract}
A paired comparison analysis is the simplest way to make comparative judgments between objects where objects may be goods, services or skills. For a set of problems, this technique helps to choose the most important problem to solve first and/or provides the solution that will be the most effective. This paper presents the theory of paired comparisons method and contributes to the paired comparisons models by developing a new model based on \emph{t}--distribution. The developed model is illustrated using a data set of citations among four famous journals of Statistics. Using Bayesian analysis, the journals are ranked as JRSS--B $\rightarrow$ Biometrika $\rightarrow$ JASA $\rightarrow$ Comm. in Stats.\\

\emph{Keywords}: paired comparisons method, ranking of objects, \emph{t}--distribution, Bayesian analysis, preference probabilities.
\newpage

\section{Introduction}
In situations where quantitative measurements for object effects are not possible, or perhaps impractical, method of rank analysis is one possible solution to analyze the outcomes of such experiments. A paired comparison (PC) analysis is probably the simplest way to make comparative judgments between objects where objects may be goods, services or skills. There is no restriction on the number and the type of objects. The technique is used primarily in experiments when weighing up the relative importance of objects. For a set of problems, this technique helps to choose the most important problem to solve first and/or provides the solution that will be the most effective. The technique compares each object against all others in pairs and sets priorities when there is a conflicting demand on the available sources. Objects are inspected pair wise by presenting them to one or more subjects to set priorities between them. To make experiment a balanced paired comparison experiment, objects are presented to more than one subjects.
\par Apart from psychology and ranking of teams and/or players, the PC methods have vast application in the field of medicine, project management, the scientific study of preferences, voting system and performance appraisal (also known as employee performance) etc.
\par Basically, there are two variants of the method of PC. The original one is due to \citet{thurstone1927} and is based on the normal distribution, called the law of comparative judgment in modern psychometric theory. The other one is known as \citet{bradley1952} model, based on the logistic distribution. \citet{thurstone1927} proposed the law of comparative judgment in which a pairwise comparison is used to scale a number of objects. \citet{bradley1952} and \citet{bradley1953} used the logistic density function to formulate the model for PC. This model proposed the preference of one object over the other. The maximum likelihood estimates are used for rating of objects. Its usefulness in sensory testing is also demonstrated. Likelihood ratio test is used to test certain hypotheses about rating parameter.
\par \cref{sec:exofpca} reviews some of the well known PC models in literature and their mathematical forms are discussed. \cref{sec:develop} elaborates the PC model building mechanism through a probabilistic approach. The mathematical form of a PC model using \emph{t}--distribution is constructed in the same Section. The formation of likelihood function is taken into account in \cref{sec:likelihood}. \cref{sec:priors} presents different prior distributions used for Bayesian analysis. \cref{sec:bag} deals with the Bayesian analysis; derivation of the posterior distributions, estimation of posterior estimates, estimation of preference and predictive probabilities, and finally the $\chi^2$ goodness of fit test. The application of the developed model is illustrated in \cref{sec:application} using a data set taken from \citet{turner2012}. This article finishes off by drawing conclusion and final remarks in \cref{sec:conclusion}.
\section{Some of the PC Models}\label{sec:exofpca}
The law of comparative judgment, proposed by \citet{thurstone1927} is applicable not only to physical objects but also for qualitative judgment. The excellence of objects, opinion on public disputed issues, sensory and taste testing, and performance appraisal are examples of this law. Thurstone suggests in 1931 that his law can also be suitable for ranking of objects. The following three points of Thurstone's (1927) model are important to be noted
\begin{itemize}
	\item A subject elicits a preference on continuous scale whenever confronted with a pair of objects
	\item The object having larger sensation value (on continuous scale) is preferred by the subject
	\item The unobserved preferences follow a normal distribution in population.
\end{itemize}
\par If $\theta_i \, \text{and} \, \sigma_i \, (i=1,2,\dots,m)$ are worth and scale parameters of object $T_i$. If $\psi_{ij}$ denotes the preference probability of $T_i$ over $T_j, (i<j,i,j=1,2,\dots,m)$ then according to \citet{thurstone1927}
\begin{align}
\psi_{ij}&=\Pr(T_i>T_j)
				  =\frac{1}{\sqrt{2\pi \left( \sigma_i^2+\sigma_j^2-2\sigma_{ij}^2 \right)}}\int \limits_{-\infty}^{(\theta_i-\theta_j)/(\sigma_i^2+\sigma_j^2-2\sigma_{ij}^2)} \exp\left({\frac{-1}{2} y_{ij}^2}\right) \text{d} y_{ij},
\end{align}
where $Y_{ij}=X_i-X_j$. Case V of \citet{thurstone1927} assumes that the correlation between objects is equal and zero (or nearly so). Hence, we are left with mean ($\theta_i-\theta_j$) and unit variance for preferences. i.e.,
\begin{align}
\nonumber
\psi_{ij}&=\Pr(T_i>T_j)
				  =\frac{1}{\sqrt{2\pi}}\int \limits_{-\infty}^{\theta_i-\theta_j} \exp\left({\frac{-1}{2} y_{ij}^2}\right) \text{d} y_{ij}, \\
				 &=\Phi(\theta_i-\theta_j),
\end{align}
where $\Phi(\cdot)$ denotes the standard normal distribution function.
\par \citet{bradley1952} modified the \citet{thurstone1927} model and used logistic distribution instead of normal distribution. The Bradley-Terry model implies that preferences follow a logistic distribution with location ($\ln\theta_i-\ln\theta_j$). So, the preference probability of $T_i$ over $T_j$ is
\begin{align}
\nonumber
\psi_{ij}&=\Pr(T_i>T_j)
				  =\frac{1}{4}\int \limits_{-\infty}^{\ln\theta_i-\ln\theta_j} \sech^2 {\left( y_{ij}/2 \right)} \text{d} y_{ij}, \\
				 &=\frac{\theta_i}{\theta_i+\theta_j}.
\end{align}
\par Following \citet{thurstone1927} and \citet{bradley1952}, \citet{abbas2009} proposed a PC model based on Cauchy distribution. For a pair of objects $(T_i,T_j)$, if $\psi_{ij}$ denotes the preference probability of $T_i$ over $T_j$ then
\begin{align}
\nonumber
\psi_{ij}&=\Pr(T_i>T_j)
				  =\frac{1}{\pi}\int \limits_{-\infty}^{\theta_i-\theta_j} \frac{1}{1+y_{ij}^2} \text{d} y_{ij}, \\
				 &=\frac{1}{2}+\frac{\tan^{-1}(\theta_i-\theta_j)}{\pi}.
\end{align}

\section{Development of a PC Model using \emph{t}--Distribution} \label{sec:develop}
This Section presents the mechanism of PC model building criteria originally proposed by \citet{thurstone1927}. In a PC experiment, subject is confronted with a series of objects such as handwriting specimens, cylindrical weights, taste testing, children' drawings or any other series of objects for ranking purpose. The first important thing is that what is going to be judged or compared. It may be quantitative or qualitative such as weight or excellence of objects. The object is preferred on the basis of some attribute of excellence. The value (or worth) assigned to objects on the basis of these attributes are called \emph{psychological continuum}. The process of reacting differently to the objects to identify the degree of excellence is called a \emph{discriminal process}. It is assumed that discriminal process is variable for a given object. i.e., the subject gives different value (or worth) of psychological continuum on successive occasions about the same pair of objects. \citet{thurstone1927} assumed that the psychological continuum is defined in such a way that frequencies of discriminal process for a given object form a normal distribution on psychological scale. It is to be noted that the scale is defined in terms of discriminal process frequency of a certain object. The difference between observed discriminal process and a model (true or ideal) discriminal process is called a \emph{discriminal deviation} on a certain occasion.
\par In this study, we assume that the frequencies of discriminal process follow a \emph{t}--distribution on psychological scale. The reason behind using \emph{t}--distribution is that it is independent of discriminal dispersion. i.e., no matter the difference between observed discriminal process and true or ideal discriminal process is small or large, the \emph{t}--distribution is independent of that.
\par Now we formally define a PC model in a probabilistic way. Suppose that there are $n$ objects $T_1,T_2,\dots,T_n$ to be compared in a balanced PC experiment. Suppose the true worth of an object $T_i\, (i=1,2,\dots,n)$ is $W_i$. When the object $T_i$ is presented to a subject in $m$ comparisons, its worth varies in each comparison and represented by a continuous random variable $Y_i\, (-\infty<Y_i<\infty)$. If $T_i$ and $T_j$ are compared, the former is preferred if $y_i>y_j$, and the latter if $y_j>y_i$. If $T_i$ is preferred over $T_j$, we write it as $T_i \rightarrow T_j$ and it's complement is written as $T_j \rightarrow T_i$. Let $Z_i=Y_i-W_i\, (i=1,2,\dots,m)$. If the $Z_i$'s are independently and identically distributed or nearly so, then $Z_i-Z_j$ must have the same distribution as $Z_j-Z_i$ i.e., the distribution of the differences $Z_i-Z_j$ is symmetric about zero. Then the preference probability that object $T_i$ is preferred over $T_j$, denoted by $\psi_{ij}$, is given in \cref{defpref}.
\begin{align}
\qquad \psi_{ij} \nonumber &=\Pr  \left (Y_i-Y_j>0\right )\\
       \nonumber &=\Pr \left (Z_i+W_i>Z_j+W_j\right )\\
			 \nonumber &=\Pr \left (Z_j-Z_i<W_i-W_j\right )\\
 \label{defpref} &= \text{H}  \left (W_i-W_j\right ),								
\end{align}
where $\text{H}(\cdot)$ is distribution function defined by $\Pr (Z_j-Z_i\leq a)=\text{H}(a)$, for any real constant $a$.
\par Whenever the preference probabilities are expressed in the form of \cref{defpref}, the random variable $Y_i$ is said to satisfy a linear model. We assume that when a series of objects is homogenous, i.e, there is a low correlation between discriminal deviations and possibly even zero, the degree of excellence which a subject perceives in one of the two objects has no influence on the degree of excellence of the other. Practically, it is assumed when a group of subjects perceives an object for comparison, the distribution of preferences is a \emph{t}--distribution with $\nu$ degrees of freedom. The location parameter of \emph{t}--distribution is ($\theta_i-\theta_j$). The probability that object $T_i$ is preferred over $T_j$, denoted by $\psi_{ij}^{(\nu)}$, is given by
\begin{align}
\nonumber \psi_{ij}^{(\nu)}&=\Pr \left (T_i-T_j>0\right ),	\quad \quad  i\neq j; \; 0<\psi_{ij}^{(\nu)}<1,    \\
\nonumber                  &=\int\limits^{\theta _i-\theta _j}_{-\infty} \frac{\left (\frac{\nu }{\nu +y^2} \right)^{\frac{\nu +1}{2}}}{\sqrt{\nu } \, \text{Beta}\left(\frac{\nu }{2},\frac{1}{2}\right)} \, \text{d} y, \\
\label{modelg}  &=\begin{cases}
\begin{aligned}
& \frac{1}{2} I_{\frac{\nu }{\nu +(\theta_i-\theta_j)^2}}\left(\frac{\nu }{2},\frac{1}{2}\right), && \qquad \theta _i-\theta _j\leq 0, \\
& \frac{1}{2} I_{\frac{(\theta_i-\theta_j)^2}{\nu +(\theta_i-\theta_j)^2}}\left(\frac{1}{2},\frac{\nu }{2}\right)+1, && \qquad \theta_i-\theta_j > 0,
\end{aligned}         \end{cases}						
\end{align}
where Beta$\left (\cdot\right )$ and $I_x\left (a,b\right )=\frac{1}{\text{Beta}\left (a,b\right )} \int^x_0 y^{a-1} \left (1-y\right )^{b-1} \, \text{d}y$, denote the beta and incomplete (regularized) beta functions respectively. In $\psi_{ij}^{(\nu)}$, $\nu$ shows that preference probability is a function of $\nu$. It is considered that the subject must differentiate between objects presented to him, i.e., tie between objects is not being considered here, therefore, $\psi_{ij}^{(\nu)}$ + $\psi_{ji}^{(\nu)}$ = 1 (where $\psi_{ji}^{(\nu)}$ is the complement of $\psi_{ij}^{(\nu)}$). Since the proposed paired comparison model in \cref{modelg} is based on \emph{t}--distribution, it is named as \emph{t}-paired comparison model (\emph{t}-PCM).
\section{Important Notations and Likelihood Function} \label{sec:likelihood}
In this Section, we define some of the notations to make an understanding of the model and for further analysis.

\begin{flushleft}
\qquad $r_{ijk} = \begin{cases}
	1, \qquad  \text{if the object $T_i$ is preferred over $T_j$ in $k$th repetition,} \\
	0, \qquad  \text{otherwise;}\,  i<j;\, i,j=1,2,\dots,n;\, k=1,2,\dots,m, \end{cases}$ \\
\qquad $r_{jik} = \begin{cases}
	1, \qquad \text{if the object\, $T_j$ is preferred over $T_i$ in $k$th repetition,} \\
	0, \qquad \text{otherwise;}\,  j<i;\, i,j=1,2,\dots,n;\, k=1,2,\dots,m, \end{cases}$ \\
\qquad  i.e., $r_{ijk} + r_{jik} = 1$, \\					
\qquad $r_{ij} = \sum\limits^{m}_{k=1} r_{ijk}$, \, the number of times object $T_i$ is preferred over $T_j$, \\
\qquad $r_i = \sum\limits^{n}_{j (\neq i)=1} r_{ij}$, \, the number of times object $T_i$ is preferred over all other objects, \\
\qquad $n_{ij}$ = the number of times object $T_i$ is compared with $T_j$. \\
\end{flushleft}

\par We observed that there are only two responses against each single comparison with fixed probability $\psi_{ij}^{(\nu)}$. It is more appropriate to call this experiment a Bernoulli experiment. i.e., The probability of $k\text{th}$ repetition of a pair of objects $(T_i,T_j)$ is
\begin{align}
P_{ijk}=\left (\psi_{ij}^{(\nu)}\right )^{r_{ijk}}\, \left (1-\psi_{ij}^{(\nu)}\right )^{r_{jik}}.
\end{align}
\par Assuming the independence in judgments by subjects, the likelihood function of the observed data $\mathbf{r}=\{n_{ij},r_{ij}\}$ for Bernoulli responses is given in \cref{likeg}
\begin{align}
\nonumber L\left (\mathbf{r};\boldsymbol{\theta}\right ) &= \prod\limits^{n}_{i(<j)=1} \prod\limits^{m}_{k=1} P_{ijk}, \\
\label{likeg}
L\left (\mathbf{r};\boldsymbol{\theta}\right ) &= \prod\limits^{n}_{i(<j)=1} \frac{n_{ij}!}{r_{ij}!\left (n_{ij}-r_{ij}\right )!}\, \left( \psi_{ij}^{(\nu)} \right)^{r_{ij}}\, \left (1-\psi_{ij}^{(\nu)}\right )^{n_{ij}-r_{ij}},
\end{align}
where $\boldsymbol{\theta}=\left (\theta_1,\theta_2,\dots,\theta_n\right )$ denotes the worth vector of $n$ objects being compared.

\section{Prior Distributions} \label{sec:priors}
The basic idea behind the Bayesian methodology is to combine information from experiment and the personal belief. The personal belief is believed to be as prior information, which may be based upon some previous experiments. The uniform and the Jeffreys priors are the most common choice for noninformative priors.

\subsection{The Uniform Prior} \label{sec:uprior}
The choice of uniform prior makes sense because it gives equal weights to all possible values of a parameter. It is used for a parameter having finite range, however, it also gives satisfactory results otherwise. We assume independent uniform priors for $\boldsymbol{\theta}=(\theta_1,\theta_2,\dots,\theta_n)$ over the interval $\left( -\infty ,\infty \right)$. The uniform joint prior is denoted by $\pi_{u}$ and, in the density kernel form, is given in \eqref{jointuprior}.
\begin{align}
\label{jointuprior}
\pi_{u} \left (\boldsymbol{\theta}\right ) \propto 1, \qquad \qquad -\infty < \theta_i < \infty.
\end{align}

\subsection{The Jeffreys Prior} \label{sec:jprior}
\citet{jeffreys1961} is first who sets the basis for constructing data priors. It is derived from Fisher's information matrix. Consider $\boldsymbol{\theta}=\left (\theta_1,\theta_2,\dots,\theta_n\right )$ to be a vector of parameters which are continuous in nature, the Jeffreys joint prior (denoted by $\pi_{d}$) is defined as
\begin{align}
\label{jointjprior}
\pi_{d} \left( \boldsymbol{\theta} \right) \propto \sqrt{\text{det} \left\{ I \, (\boldsymbol{\theta}) \right\}},
\end{align}
where ``det'' denotes the determinant and $I \, (\boldsymbol{\theta})$ is Fisher's information matrix and is defined as $I \, (\boldsymbol{\theta})=-E \left\{ \tfrac{\partial ^2 \, \text{ln} \, L \, ( \textbf{x} ; \boldsymbol{\theta})}{\partial \, \boldsymbol{\theta} ^2} \right\}$. Where expectation is with respect to data, $\textbf{x}= (x_1,x_2,\dots,x_n)$, and $L \, ( \textbf{x} ; \boldsymbol{\theta})$ denotes the likelihood function.

\section{Bayesian Analysis of \emph{t}--PCM}\label{sec:bag}
The initial step in Bayesian analysis is the utilization of prior knowledge about parameters under consideration. The joint prior, we generally denote by $\pi(\boldsymbol{\theta})$ (given in \eqref{jointuprior} and \eqref{jointjprior}), is incorporated with the likelihood function in \cref{likeg} to yield joint posterior. The joint posterior distribution of $\boldsymbol{\theta}$ given data is obtained as
\begin{align}
\nonumber p \left( \boldsymbol{\theta} | \mathbf{r} \right) &\propto \prod\limits^{n}_{i(<j)=1} \left( \psi_{ij}^{(\nu)} \right)^{r_{ij}}\, \left( 1-\psi_{ij}^{(\nu)} \right)^{n_{ij}-r_{ij}} \pi \left( \boldsymbol{\theta} \right), \\
\label{jpostg} p \left( \boldsymbol{\theta} | \mathbf{r} \right) &= \frac{1}{C} \prod\limits^{n}_{i(<j)=1} \left( \psi_{ij}^{(\nu)} \right)^{r_{ij}}\, \left( 1-\psi_{ij}^{(\nu)} \right)^{n_{ij}-r_{ij}} \pi \left( \boldsymbol{\theta} \right), \quad -\infty < \theta_i < \infty,
\end{align}
where $\mathbf{r}=(n_{ij},r_{ij})$ is data and $C$ is the constant of proportionality and is given by
\begin{align*}
C &= \int\limits_{\boldsymbol{\theta}} \, \prod\limits^{n}_{i(<j)=1} \left( \psi_{ij}^{(\nu)} \right)^{r_{ij}}\, \left( 1-\psi_{ij}^{(\nu)} \right)^{n_{ij}-r_{ij}} \pi \left( \boldsymbol{\theta} \right) \text{d} \boldsymbol{\theta}.
\end{align*}

\subsection{Marginal Posterior Distributions} \label{sec:mpostgi}
The marginal posterior distribution of parameters (or worth parameters) given data are obtained by integrating \cref{jpostg} over nuisance parameters. The marginal posterior distribution of $\theta_i \, (i=1,2,\dots,n)$ given data is
\begin{align}
\label{mpostgi} p \left( \theta_i | \mathbf{r} \right) &= \frac{1}{C} \int\limits_{\boldsymbol{\theta^{'}}} \prod\limits^{n}_{i(<j)=1} \left( \psi_{ij}^{(\nu)} \right)^{r_{ij}}\, \left( 1-\psi_{ij}^{(\nu)} \right)^{n_{ij}-r_{ij}} \pi \left( \boldsymbol{\theta} \right) \text{d} \boldsymbol{\theta^{'}}, \quad -\infty < \theta_i < \infty,
\end{align}
where $\boldsymbol{\theta^{'}}$ is such that $\boldsymbol{\theta^{'}} \cup \theta_i = \boldsymbol{\theta}$ and $\boldsymbol{\theta^{'}} \cap \theta_i = \boldsymbol{\Phi}$. The posterior estimates (e.g., posterior means and posterior modes) are used for calculation of preference probabilities.

\subsection{Posterior Estimates} \label{sec:postestimatesg}
The posterior mean of $i\text{th}\, (i=1,2,\dots,n)$ object is obtained by taking expectation of $\theta_i$ with respect to its marginal posterior distribution. i.e.,
\begin{align}
 \label{expgi}
E\left (\theta_i | \mathbf{r}\right ) &= \int\limits_{-\infty}^{\infty} \theta_i \, p\left (\theta_i | \mathbf{r}\right ) \text{d} \theta_i \, ,
\end{align}
where $p(\theta_i | \mathbf{r})$ is given in \cref{mpostgi}. Due to the complicated expressions of marginal posterior distributions and hence posterior means, we use quadrature method to obtain posterior estimates. The quadrature method is simple and is similar to the weighted mean. According to this method, the value of $\int_a^b {g(\theta)} \, \text{d} \theta$ can be approximated by evaluating $g(\theta)$ as a weighted sum of function at specified small intervals over the whole range of $\theta$ i.e., $\int_a^b {g(\theta)} \, \text{d} \theta \approx\sum_{i=1}^{n} w_i \, g(\theta_i)$, where $w_i$ are the weights to be used $a$ through $b$ and $n$ is the total no. of interval points. The two-dimensional integration is approximated by $\int_c^d \int_a^b {g(\theta_i,\theta_j)} \, \text{d} \theta_i \, \text{d} \theta_j \approx\sum_{j=1}^{m} \sum_{i=1}^{n} w_i \, w_j \, g(\theta_i,\theta_j)$, and higher dimensional integration is approximated in a similar manner.
\par The posterior modes in uniform priors are derived using first derivatives of log-likelihood function whereas joint posterior distribution is maximized in Jeffreys priors subject to the constraint $\sum_{i=1}^{n}\theta_i=0$.

\subsection{Preference Probabilities} \label{sec:prefprobg}
Preference probability shows the expected chance of preferring one object over the other in a pairwise comparison. To find the preference probabilities, we substitute the posterior estimates of objects in \cref{modelg} i.e., $\psi_{ij}^{(\nu)}$. The posterior means and posterior modes are obtained to estimate preference probabilities.

\subsection{Predictive Probabilities} \label{sec:predg}
The predictive probabilities are obtained by multiplying \cref{modelg} with joint posterior distribution in \cref{jpostg} and then integrating over all the worth parameters. If $\eta_{lm}^{(\nu)} \, \big\{l(<m)=1,2,\dots,n\big\}$ denotes the predictive probability when objects $T_l$ and $T_m$ would be compared $n_{lm}$ times, then
\begin{align}
\label{predprobsg}
\eta_{lm}^{(\nu)} &= \int\limits_{\boldsymbol{\theta}}^{} \psi_{lm}^{(\nu)} \, p \left( \boldsymbol{\theta} | \mathbf{r} \right) \, \text{d} \boldsymbol{\theta}, \qquad l<m.
\end{align}

\subsection{Goodness of Fit Test} \label{sec:goffitg}
The appropriateness of the \emph{t}-PCM (given in \cref{modelg}) with observed data (\cref{tab:obsdata}) is checked using $\chi^2$ goodness of fit test. The procedure of $\chi^2$ is to calculate the theoretical (expected) frequencies and compare them with observed frequencies to check the differences. If the differences appear to be small then the model is considered to give good fit of data otherwise not. The following hypotheses are formulated.
\begin{flushleft}
$H_0$ \, : \, The model is appropriate for $\boldsymbol{\theta}=\boldsymbol{\theta}_{0}$, \\
$H_1$ \, : \, The model is not appropriate for any value of $\boldsymbol{\theta}$.
\end{flushleft}
\par Following (\citet{abbas2009}), if $H_0$ is true, the $\chi^2$-statistic to be used to check the appropriateness of the model is given by
\begin{align}
\label{chiformula}
\chi^2 &= \sum\limits_{i(<j)=1}^{n}
	\left \{ \frac{\left (r_{ij}-\widehat{r}_{ij}\right )^{2}}{\widehat{r}_{ij}}+\frac{\left (r_{ji}-\widehat{r}_{ji}\right )^{2}}{\widehat{r}_{ji}}
	\right \},
\end{align}
with $\tfrac{(n-1)(n-2)}{2}$ degrees of freedom ($df$). And where the \^ \, above observed frequencies ($r_{ij}$) shows that the frequencies are theoretical (expected). The expected frequencies are obtained by substituting posterior estimates in $\widehat{r}_{ij} = n_{ij} \, \psi_{ij}^{(\nu)}$.

\section{Application of \emph{t}-PCM using Real Data}\label{sec:application}
The analyses of \emph{t}--PCM are carried out using the data set taken from \citet{turner2012}. The data set in \cref{tab:obsdata} shows the counts of citations among four famous journals of Statistics during 1987--1989. We use the aliases 1, 2, 3 and 4 for Biometrika, Communication in Statistics (Comm. in Stats.), Journal of American Statistical Association (JASA) and Journal of Royal Statistical Society--Series B (JRSS--B), respectively. The leading column and the leading row show the cited journals and the citing journals, respectively. Thus, for instance, Biometrika is cited 730 times by Comm. in Stats. The preference probabilities are based on the number of respective preferences. The value 0.95675 indicates that preference probability of Biometrika over Comm. in Stats. is 0.95675.
\par
\begin{figure}[h]
	\centering
	\includegraphics[width=0.8\textwidth]{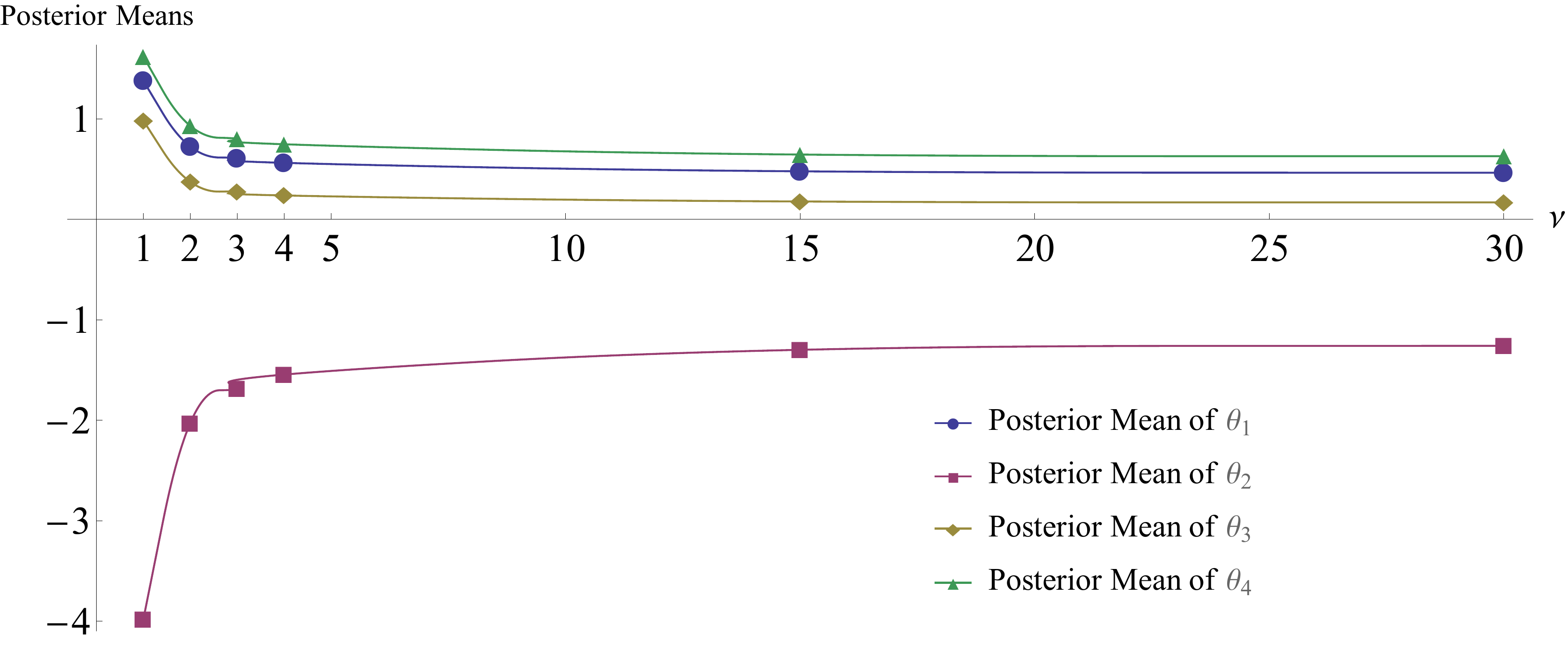}
	\caption{Posterior Means of Worth Parameters using Uniform Prior}
	\label{fig:upt4jointmeans}
\end{figure}

\begin{figure}[h]
	\centering
	\includegraphics[width=0.8\textwidth]{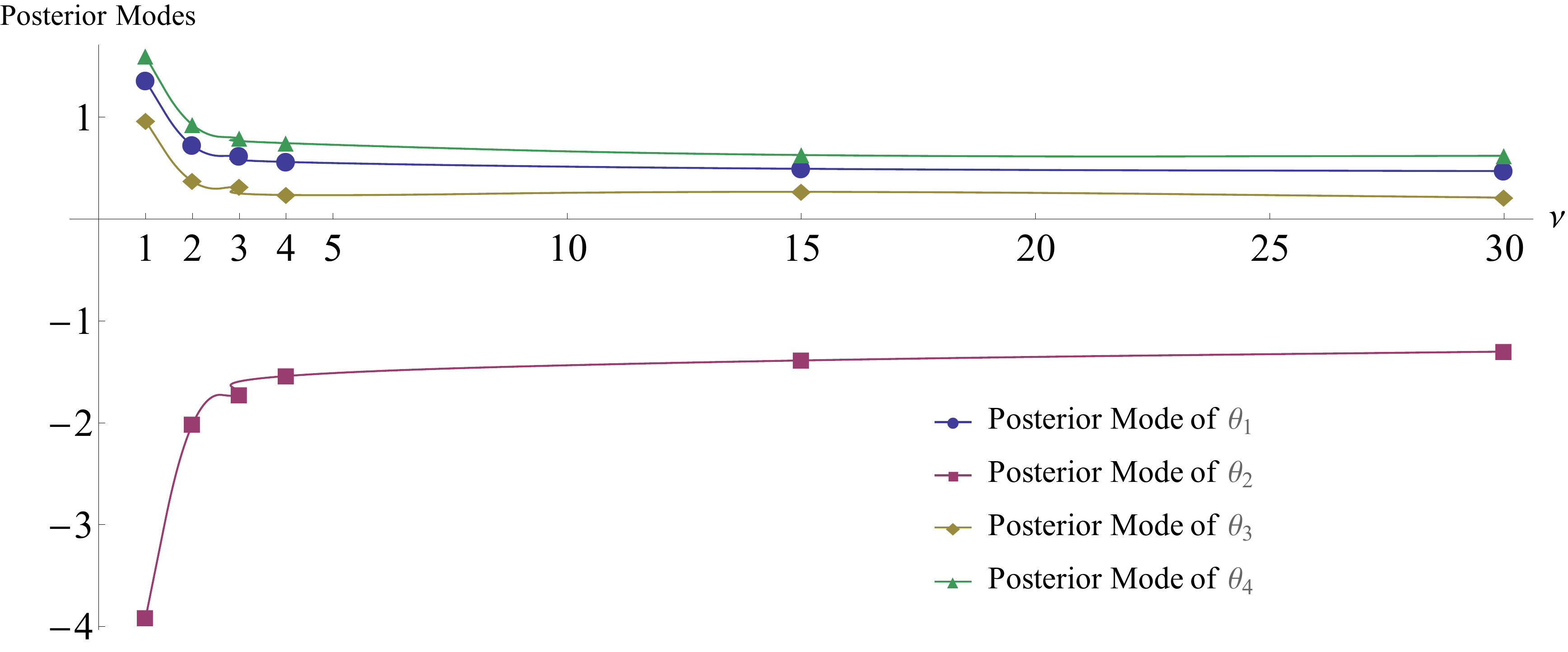}
	\caption{Posterior Modes of Worth Parameters using Uniform Prior}
	\label{fig:upt4jointmodes}
\end{figure}

\begin{figure}[h]
	\centering
	\includegraphics[width=0.8\textwidth]{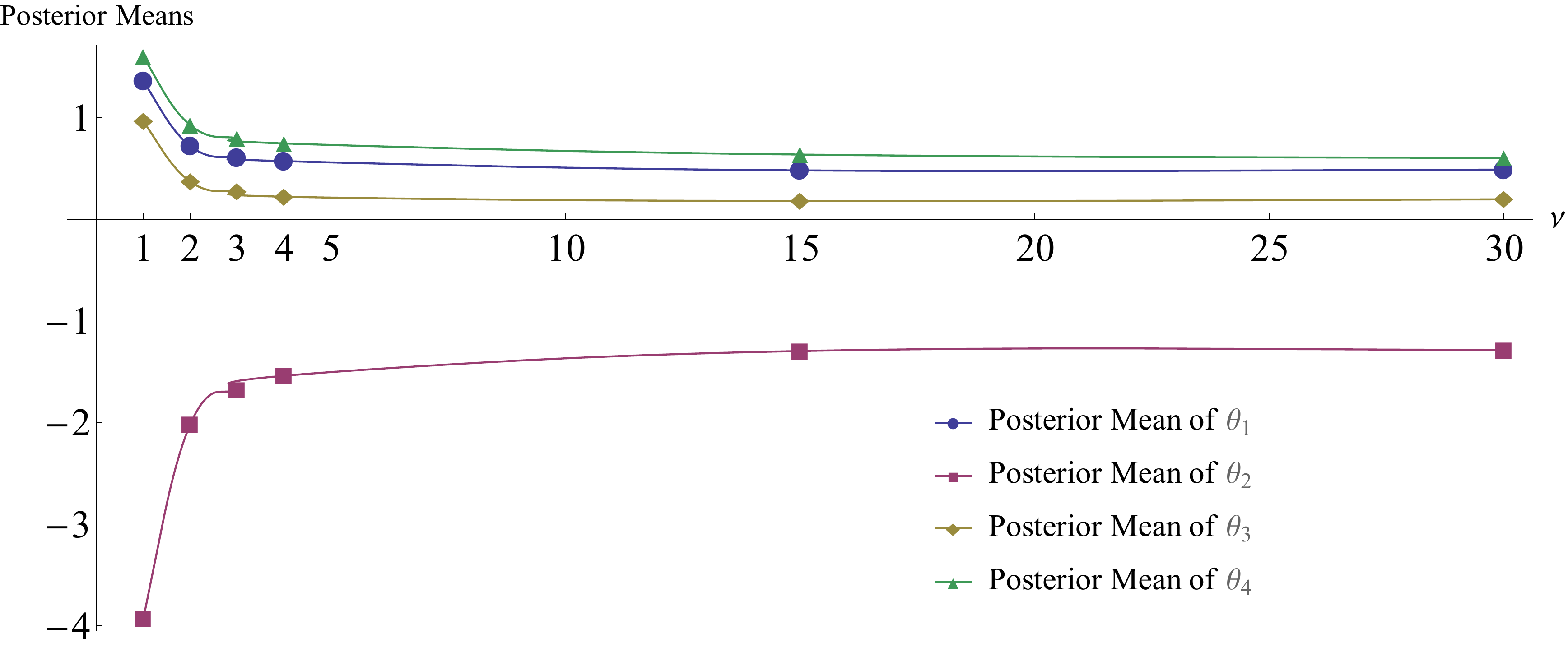}
	\caption{Posterior Means of Worth Parameters using Jeffreys Prior}
	\label{fig:jpt4jointmeans}
\end{figure}

\begin{figure}[h]
	\centering
	\includegraphics[width=0.8\textwidth]{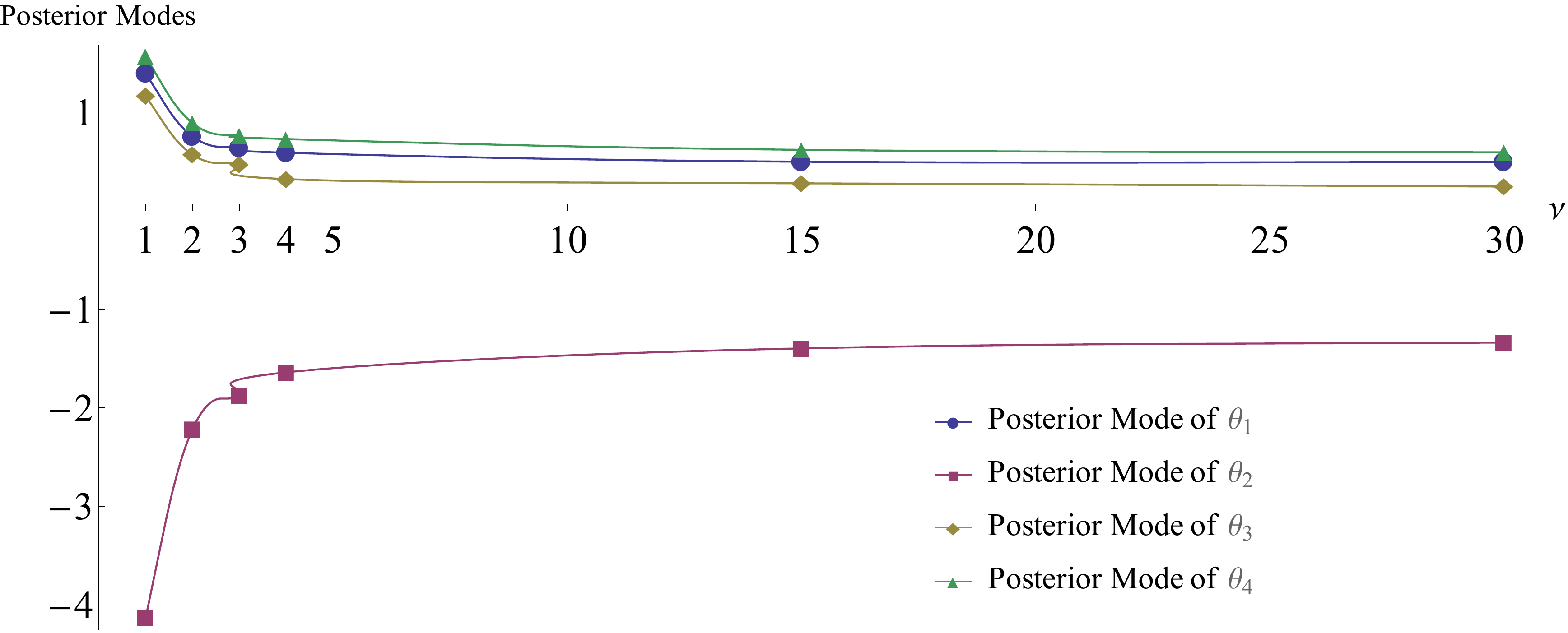}
	\caption{Posterior Modes of Worth Parameters using Jeffreys Prior}
	\label{fig:jpt4jointmodes}
\end{figure}

 First of all, we shall discuss the graphical behavior of posterior estimates. The joint plotting of the posterior estimates give a better look how the worth behave with varying $\nu$. The posterior estimates using uniform and Jeffreys priors are plotted in \cref{fig:upt4jointmeans,fig:upt4jointmodes,fig:jpt4jointmeans,fig:jpt4jointmodes}. In \cref{fig:upt4jointmeans}, the line on top shows the posterior mean of $\theta_4$ for different values of $\nu$. These figures show that $\theta_4$ is on the top of all, ranked as first and $\theta_1$ is ranked as second. The parameter $\theta_3$ is above the $x$-axis but below the line of $\theta_1$, it is ranked as third and $\theta_2$ ranked as fourth. Therefore, the ranking based on posterior estimates is JRSS--B $\rightarrow$ Biometrika $\rightarrow$ JASA $\rightarrow$ Comm. in Stats.
\par The \cref{tab:meansmodesuj4} shows posterior estimates of worth parameters. It can be readily observed that there is a minor difference in posterior means and posterior modes. The posterior estimates of $\theta_4$ are highest among all posterior estimates (e.g., posterior mean of $\theta_4=1.62080$ for $\nu=1$ is highest than $\theta_1=1.37908$, $\theta_2=-3.98254$ and $\theta_3=0.98266$). The posterior estimate of Biometrika is greater than JASA. The Comm. in Stats. has minimum posterior estimate. So, the ranking based on posterior estimates is JRSS--B $\rightarrow$ Biometrika $\rightarrow$ JASA $\rightarrow$ Comm. in Stats. As can be seen from the table that other values of $\nu$ agree with this ranking. Moreover, posterior estimates seem to be less changing their behavior for $\nu\geq4$, as shown by \cref{fig:upt4jointmeans,fig:upt4jointmodes,fig:jpt4jointmeans,fig:jpt4jointmodes}.
\par The preference probabilities obtained using posterior means and posterior modes are displayed in \cref{tab:prefprobsmeansuj4,tab:prefprobsmodesuj4}, respectively. The preference probabilities calculated using either prior are almost same. The preference probabilities of $\theta_1$, $\theta_3$ and $\theta_4$ over $\theta_2$ are all greater than 0.90. The preference probabilities of $\theta_1$ and $\theta_4$ over $\theta_3$ are greater than 0.60 implies JASA is ranked as third. The JRSS--B has preference probability greater than Biometrika, so, JRSS--B is ranked as first. Thus, the ranks derived from preference probabilities are JRSS--B $\rightarrow$ Biometrika $\rightarrow$ JASA $\rightarrow$ Comm. in Stats. The ranking is same for all values of $\nu$.
\par The \cref{tab:predprobsuj4} shows the predictive probabilities for a single future preference. The predictive probabilities calculated using uniform priors do not vary much from that obtained using Jeffreys priors. The predictive probabilities of citations for $\theta_2$ are less than 0.10 predicting that Comm. in Stats. will be cited less among these journals. The worth parameter $\theta_4$ has all the predictive probabilities greater than 0.50 implying JRSS--B will be cited most.
\par The $\chi^2$ goodness of fit test reveals that \textit{t}-PCM fits best for $\nu=2$, 3 and 4 with a $p$-value of greater than 0.15.

\section{Conclusion and Remarks}\label{sec:conclusion}
A new model for comparative judgment is developed following \citet{thurstone1927}. This new model (\emph{t}--PCM) is based on \emph{t}--distribution. It will be used for ranking of objects where quantitative measurements for object effects are not possible, or perhaps impractical such as handwriting specimens, taste testing, children' drawing etc.
\par For estimation and prediction of worth parameters of the model, uniform and Jeffreys priors are incorporated using Bayesian statistics. The Bayesian analyses constitute the estimation of posterior estimates, preference probabilities, predictive probabilities and expected citations. The appropriateness of model is checked using $\chi^2$ goodness of fit test. All the analyses are carried out using a data set showing the counts of citations among four famous journals of Statistics.
The examination of graphs and estimates of probabilities reveals that the journals are ranked as; JRSS--B stands first as it has highest preference and predictive probabilities; following the JRSS--B, Biometrika is at second rank; JASA got third position and Comm. in Stats. ranked last. 
\par The objective of choosing two noninformative priors is to explore the effect of a particular prior on preference probabilities and hence on ranking. In general, the comparison is concluded with the discussion that uniform and Jeffreys priors both are equally efficient. Both priors can be used alternatively however working with Jeffreys prior is somewhat tedious than the simpler uniform prior.

\bibliographystyle{apalike}
\bibliography{apm}
\newpage
\appendix
\section{Appendix}

\begin{table}[h]
  \centering
  \caption{Observed Data of Citations Among Four Journals: 1987--1989}
\scalebox{0.90}{
    \begin{tabular}{cccccc}
    \toprule
		& & \multicolumn{4}{c}{Citing} \\
    & Cited & 1     & 2     & 3     & 4 \\
    \cmidrule(r){2-6}
    \multirow{4}{*}{No. of Preferences}
					& 1     & -     & 730   & 498   & 221 \\
          & 2     & 33    & -     & 68    & 17 \\
          & 3     & 320   & 813   & -     & 142 \\
          & 4     & 284   & 276   & 325   & - \\
		\cmidrule(r){2-6}			
    \multirow{4}{*}{Preference Probabilities}
					& 1     & -     	& 0.95675 & 0.60880 & 0.43762 \\
          & 2     & 0.04325 & -     	& 0.07718 & 0.05802 \\
          & 3     & 0.39119 & 0.92281 & -     	& 0.30407 \\
          & 4     & 0.56238 & 0.94198 & 0.69593 & - \\
    \bottomrule
    \end{tabular}}%
  \label{tab:obsdata}%
\end{table}%

\begin{table}[h]
  \centering
  \caption{Posterior Estimates of Worth Parameters}
\scalebox{0.90}{
    \begin{tabular}{ccccccccc}
    \toprule
    \multirow{1}[0]{*}{} & \multicolumn{4}{c}{Posterior Means} & \multicolumn{4}{c}{Posterior Modes} \\
    \midrule
     $\nu$   & $\theta_1$ & $\theta_2$ & $\theta_3$ & $\theta_4$ & $\theta_1$ & $\theta_2$ & $\theta_3$ & $\theta_4$ \\
		\midrule
		\multicolumn{9}{c}{Uniform Prior}\\
		\midrule
     1     & 1.37908	& -3.98254	& 0.98266	& 1.62080	& 1.35722 & -3.91721 & 0.96303 & 1.59696 \\
     2     & 0.72445	& -2.02793	& 0.37334	& 0.93014	& 0.72057 & -2.01634 & 0.37032 & 0.92545 \\
     3     & 0.60862	& -1.68375	& 0.27432	& 0.80081	& 0.61602 & -1.72593 & 0.31650 & 0.79341 \\
     4     & 0.56167	& -1.54542	& 0.23708	& 0.74667	& 0.56037 & -1.54127 & 0.23599 & 0.74491 \\
     15    & 0.47751	& -1.29895	& 0.17721	& 0.64423	& 0.49328 & -1.38834 & 0.26660 & 0.62846 \\
     30    & 0.46338	& -1.25942	& 0.16802	& 0.62802	& 0.47079 & -1.30160 & 0.21020 & 0.62061 \\
    \midrule
		\multicolumn{9}{c}{Jeffreys Prior}\\
		\midrule
     1     & 1.36279	& -3.93377	& 0.96793	& 1.60305 & 1.39773	& -4.13377	& 1.16793	& 1.56811 \\
     2     & 0.72135	& -2.01865	& 0.37092	& 0.92639 & 0.75629 & -2.21865 & 0.57092 & 0.89145 \\
     3     & 0.60683	& -1.67871	& 0.27311	& 0.79878 & 0.64177 & -1.87871 & 0.47311 & 0.76384 \\
     4     & 0.57200	& -1.54016	& 0.22195	& 0.74620 & 0.58947 & -1.64016 & 0.32195 & 0.72873 \\
     15    & 0.48099	& -1.29688	& 0.17921	& 0.63668 & 0.49846 & -1.39688 & 0.27921 & 0.61921 \\
     30    & 0.48840	& -1.28792	& 0.19643	& 0.60309 & 0.49713 & -1.33792 & 0.24643 & 0.59435 \\
    \bottomrule
    \end{tabular}}%
  \label{tab:meansmodesuj4}%
\end{table}%



\begin{table}[h]
  \centering
  \caption{Preference Probabilities using Posterior Means}
\scalebox{0.75}{
    \begin{tabular}{cccccccccc}
    \toprule
          &       & \multicolumn{4}{c}{Uniform Prior} & \multicolumn{4}{c}{Jeffreys Prior} \\
    \midrule
    $\nu$ &  & 1 & 2 & 3 & 4 & 1 & 2 & 3 & 4 \\
		\midrule
    \multirow{4}[0]{*}{1}& 1 &  -	& 0.94130 & 0.62013 & 0.42450 &  -     & 0.94060 & 0.61970 & 0.42494 \\
    & 2 & 0.05869 & -     & 0.06326 & 0.05621 & 0.05940 & -     & 0.06406 & 0.05688 \\
    & 3 & 0.37986 & 0.93674 & -     & 0.31920 & 0.38030 & 0.93594 & -     & 0.31989 \\
    & 4 & 0.57549 & 0.94378 & 0.68079 & -     & 0.57506 & 0.94312 & 0.68011 & -     \\ 
		\midrule
    \multirow{4}[0]{*}{2}& 1 & -     & 0.94473 & 0.62048 & 0.42803 & -     & 0.94431 & 0.62026 & 0.42826 \\
    & 2 & 0.05527 & -     & 0.06917 & 0.04891 & 0.05569 & -     & 0.06971 & 0.04927 \\
    & 3 & 0.37952 & 0.93083 & -     & 0.31683 & 0.37974 & 0.93029 & -     & 0.31721 \\
    & 4 & 0.57197 & 0.95110 & 0.68317 & -     & 0.57174 & 0.95073 & 0.68279 & -     \\
		\midrule
    \multirow{4}[0]{*}{3}& 1 & -     & 0.94714 & 0.61992 & 0.42993 & -     & 0.94680 & 0.61972 & 0.43002 \\
    & 2 & 0.05286 & -     & 0.07256 & 0.04445 & 0.05320 & -     & 0.07301 & 0.04473 \\
    & 3 & 0.38008 & 0.92744 & -     & 0.31750 & 0.38028 & 0.92699 & -     & 0.31775 \\
    & 4 & 0.57007 & 0.95554 & 0.68250 & -     & 0.56998 & 0.95527 & 0.68225 & -     \\
		\midrule
    \multirow{4}[0]{*}{4}& 1 & -     & 0.94859 & 0.61912 & 0.43112 & -     & 0.94888 & 0.62802 & 0.43509 \\
    & 2 & 0.05141 & -     & 0.07462 & 0.04183 & 0.05112 & -     & 0.07642 & 0.04210 \\
    & 3 & 0.38088 & 0.92537 & -     & 0.31858 & 0.37198 & 0.92358 & -     & 0.31391 \\
    & 4 & 0.56888 & 0.95817 & 0.68142 & -     & 0.56492 & 0.95790 & 0.68610 & -     \\
		\midrule
    \multirow{4}[0]{*}{15}& 1 & -     & 0.95203 & 0.61596 & 0.43491 & -     & 0.95215 & 0.61652 & 0.43918 \\
    & 2 & 0.04797 & -     & 0.08029 & 0.03550 & 0.04785 & -     & 0.08030 & 0.03613 \\
    & 3 & 0.38404 & 0.91971 & -     & 0.32360 & 0.38348 & 0.91970 & -     & 0.32694 \\
    & 4 & 0.56509 & 0.96450 & 0.67640 & -     & 0.56082 & 0.96387 & 0.67306 & -     \\
		\midrule
    \multirow{4}[0]{*}{30}& 1 & -     & 0.95239 & 0.61512 & 0.43517 & -     & 0.95709 & 0.61384 & 0.45473 \\
    & 2 & 0.04761 & -     & 0.08189 & 0.03440 & 0.04291 & -     & 0.07407 & 0.03416 \\
    & 3 & 0.38488 & 0.91811 & -     & 0.32442 & 0.38616 & 0.92593 & -     & 0.34357 \\
    & 4 & 0.56483 & 0.96559 & 0.67558 & -     & 0.54527 & 0.96584 & 0.65643 & -     \\
		\bottomrule
    \end{tabular}}%
  \label{tab:prefprobsmeansuj4}%
\end{table}%

\begin{table}[h]
  \centering
  \caption{Preference Probabilities using Posterior Modes}
\scalebox{0.75}{
    \begin{tabular}{cccccccccc}
    \toprule
          &       & \multicolumn{4}{c}{Uniform Prior} & \multicolumn{4}{c}{Jeffreys Prior} \\
    \midrule
    $\nu$ &  & 1 & 2 & 3 & 4 & 1 & 2 & 3 & 4 \\
		\midrule
    \multirow{4}[0]{*}{1}& 1 &  -	& 0.94130 & 0.62013 & 0.42450 & -     & 0.94307 & 0.57189 & 0.44628 \\
    & 2 & 0.05869 & -     & 0.06326 & 0.05621 & 0.05693 & -     & 0.05934 & 0.05526 \\
    & 3 & 0.37986 & 0.93674 & -     & 0.31920 & 0.42811 & 0.94066 & -     & 0.37883 \\
    & 4 & 0.57549 & 0.94378 & 0.68079 & -     & 0.55372 & 0.94474 & 0.62117 & - \\
		\midrule
    \multirow{4}[0]{*}{2}& 1 & -     & 0.94473 & 0.62048 & 0.42803 & -     & 0.95157 & 0.56498 & 0.45243 \\
    & 2 & 0.05527 & -     & 0.06917 & 0.04891 & 0.04843 & -     & 0.05404 & 0.04485 \\
    & 3 & 0.37952 & 0.93083 & -     & 0.31683 & 0.43502 & 0.94596 & -     & 0.38948 \\
    & 4 & 0.57197 & 0.95110 & 0.68317 & -     & 0.54757 & 0.95515 & 0.61052 & - \\
		\midrule
    \multirow{4}[0]{*}{3}& 1 & -     & 0.94714 & 0.61992 & 0.42993 & -     & 0.95693 & 0.56160 & 0.45528 \\
    & 2 & 0.05286 & -     & 0.07256 & 0.04445 & 0.04307 & -     & 0.05007 & 0.03874 \\
    & 3 & 0.38008 & 0.92744 & -     & 0.31750 & 0.43840 & 0.94993 & -     & 0.39510 \\
    & 4 & 0.57007 & 0.95554 & 0.68250 & -     & 0.54472 & 0.96126 & 0.60490 & - \\
		\midrule
    \multirow{4}[0]{*}{4}& 1 & -     & 0.94859 & 0.61912 & 0.43112 & -     & 0.95518 & 0.59885 & 0.44799 \\
    & 2 & 0.05141 & -     & 0.07462 & 0.04183 & 0.04482 & -     & 0.06063 & 0.03846 \\
    & 3 & 0.38088 & 0.92537 & -     & 0.31858 & 0.40115 & 0.93937 & -     & 0.35250 \\
    & 4 & 0.56888 & 0.95817 & 0.68142 & -     & 0.55201 & 0.96154 & 0.64750 & - \\
		\midrule
    \multirow{4}[0]{*}{15}& 1 & -     & 0.95203 & 0.61596 & 0.43491 & -     & 0.96126 & 0.58530 & 0.45275 \\
    & 2 & 0.04797 & -     & 0.08029 & 0.03550 & 0.03874 & -     & 0.05722 & 0.03103 \\
    & 3 & 0.38404 & 0.91971 & -     & 0.32360 & 0.41470 & 0.94278 & -     & 0.36928 \\
    & 4 & 0.56509 & 0.96450 & 0.67640 & -     & 0.54725 & 0.96897 & 0.63072 & - \\
		\midrule
    \multirow{4}[0]{*}{30}& 1 & -     & 0.95239 & 0.61512 & 0.43517 & -     & 0.96178 & 0.59812 & 0.46160 \\
    & 2 & 0.04761 & -     & 0.08189 & 0.03440 & 0.03822 & -     & 0.06180 & 0.03141 \\
    & 3 & 0.38488 & 0.91811 & -     & 0.32442 & 0.40188 & 0.93820 & -     & 0.36517 \\
    & 4 & 0.56483 & 0.96559 & 0.67558 & -     & 0.53840 & 0.96859 & 0.63484 & - \\
    \bottomrule
    \end{tabular}}%
  \label{tab:prefprobsmodesuj4}%
\end{table}%

\begin{table}[h]
  \centering
  \caption{Predictive Probabilities of Worth Parameters}
\scalebox{0.70}{
    \begin{tabular}{cccccccccc}
    \toprule
          &       & \multicolumn{4}{c}{Uniform Prior} & \multicolumn{4}{c}{Jeffreys Prior} \\
    \midrule
    $\nu$ & & 1 & 2 & 3 & 4 & 1 & 2 & 3 & 4 \\
		\midrule
		\multirow{4}{*}{1} & 1     & -     & 0.93766 & 0.62021 & 0.42657 &  -     & 0.93900 & 0.62115 & 0.43630 \\
          & 2     & 0.06234 & -     & 0.06754 & 0.05961 &   0.06100 & -     & 0.06602 & 0.05872 \\
          & 3     & 0.37979 & 0.93246 & -     & 0.32073 & 0.37885 & 0.93398 & -     & 0.32731 \\
					& 4     & 0.57343 & 0.94040 & 0.67927 & -  & 0.56370 & 0.94128 & 0.67269 & - \\
		\midrule
    \multirow{4}{*}{2} & 1     & -     & 0.94418 & 0.61833 & 0.42792 & -     & 0.95220 & 0.63589 & 0.43011 \\
          & 2     & 0.05582 & -     & 0.06969 & 0.04935 & 0.04780 & -     & 0.06084 & 0.04274 \\
          & 3     & 0.38167 & 0.93031 & -     & 0.31868 & 0.36411 & 0.93916 & -     & 0.30499 \\
					& 4     & 0.57208 & 0.95065 & 0.68132 & -  & 0.56989 & 0.95726 & 0.69501 & - \\
		\midrule
    \multirow{4}{*}{3} & 1     & -     & 0.94722 & 0.61304 & 0.42825 & -     & 0.94823 & 0.63276 & 0.42015 \\
          & 2     & 0.05278 & -     & 0.07115 & 0.04423 & 0.05177 & -     & 0.07682 & 0.04413 \\
          & 3     & 0.38696 & 0.92886 & -     & 0.32230 & 0.36724 & 0.92318 & -     & 0.38721 \\
					& 4     & 0.57175 & 0.95577 & 0.67770 & -  & 0.57985 & 0.95587 & 0.61279 & - \\
		\midrule
    \multirow{4}{*}{4} & 1     & -     & 0.94834 & 0.61901 & 0.43121 & -     & 0.94837 & 0.62780 & 0.43526 \\
          & 2     & 0.05166 & -     & 0.07496 & 0.04205 & 0.05163 & -     & 0.07709 & 0.04251 \\
          & 3     & 0.38101 & 0.92504 & -     & 0.31882 & 0.37220 & 0.92291 & -     & 0.32112 \\
					& 4     & 0.56879 & 0.95795 & 0.68118 & -  & 0.56474 & 0.95749 & 0.67888 & - \\
		\midrule
		\multirow{4}{*}{15} & 1     & -     & 0.95474 & 0.61576 & 0.45409 & -     & 0.93488 & 0.57814 & 0.42231 \\
          & 2     & 0.04526 & -     & 0.07597 & 0.03677 & 0.06512 & -     & 0.09085 & 0.04596 \\
          & 3     & 0.38424 & 0.92403 & -     & 0.34135 & 0.42186 & 0.90915 & -     & 0.34739 \\
					& 4     & 0.54591 & 0.96323 & 0.65865 & -  & 0.57769 & 0.95404 & 0.65261 & - \\
		\midrule
    \multirow{4}{*}{30} & 1     & -     & 0.95701 & 0.61407 & 0.45510 & -     & 0.93996 & 0.57859 & 0.42142 \\
          & 2     & 0.04301 & -     & 0.07407 & 0.03417 & 0.06004 & -     & 0.08589 & 0.04096 \\
          & 3     & 0.38594 & 0.92594 & -     & 0.34366 & 0.42141 & 0.91411 & -     & 0.34510 \\
					& 4     & 0.54490 & 0.96583 & 0.65634 & -  & 0.57858 & 0.95904 & 0.65490 & - \\
    \bottomrule
    \end{tabular}}%
  \label{tab:predprobsuj4}%
\end{table}%

\begin{table}[h]
  \centering
  \caption{Chi-Square Goodness of Fit}
\scalebox{0.70}{
    \begin{tabular}{cccccccccccc}
    \toprule
					&       & \multicolumn{4}{c}{Expected Citations} & & \multicolumn{4}{c}{Expected Citations} & \\
          &       & \multicolumn{4}{c}{(Uniform Prior)} & $\chi^2$-statistic & \multicolumn{4}{c}{(Jeffreys Prior)} & $\chi^2$-statistic \\
    \midrule
    $\nu$ & & 1 & 2 & 3 & 4 & ($p$-value) & 1 & 2 & 3 & 4 & ($p$-value) \\
		\midrule
		\multirow{4}[0]{*}{1} & 1     & -     & 718   & 507 & 214 &  & - & 718   & 507   & 215 &  \\
          & 2     & 45    & -     & 56 & 16 & 7.51316 & 45    & -     & 56    & 17 & 7.34127 \\
          & 3     & 311   & 825   & - & 149 & (0.05722) & 311   & 825   & -     & 149 & (0.06178) \\
          & 4     & 291   & 277   & 318 & - & & 290   & 276   & 318   & - & \\
		\midrule
		\multirow{4}[0]{*}{2} & 1     & -     & 721   & 508 & 216 &  & -     & 720   & 503   & 222 &  \\
          & 2     & 42    & -     & 61 & 14 & 4.65684 & 43    & -     & 62    & 15 & 5.13922 \\
          & 3     & 310   & 820   & - & 148 & (0.19872) & 315   & 819   & -     & 155 & (0.16188) \\
          & 4     & 289   & 279   & 319 & - & & 283   & 278   & 312   & - & \\
		\midrule
		\multirow{4}[0]{*}{3} & 1     & -     & 723   & 507 & 217 &  & -     & 723   & 502   & 216 &  \\
          & 2     & 40    & -     & 64 & 13 & 3.75587 & 40    & -     & 63    & 13 & 3.92138\\
          & 3     & 311   & 817   & - & 148 & (0.28906) & 316   & 818   & -     & 150 & (0.27008) \\
          & 4     & 288   & 280   & 319 & - & & 289   & 280   & 317   & - & \\
		\midrule
		\multirow{4}[0]{*}{4} & 1     & -     & 724   & 506 & 218 &  & -     & 724   & 514   & 220 &  \\
          & 2     & 39    & -     & 66 & 12 & 4.09782 & 39    & -     & 67    & 12 & 4.75766\\
          & 3     & 312   & 815   & - & 149 & (0.25109) & 304   & 814   & -     & 147 & (0.19043) \\
          & 4     & 287   & 281   & 318 & - & & 285   & 281   & 320   & - & \\
		\midrule
		\multirow{4}[0]{*}{15} & 1     & -     & 726   & 504 & 220 &  & -     & 726   & 504   & 222 &  \\
          & 2     & 37    & -     & 71 & 10 & 6.65237 & 37    & -     & 71    & 11 & 5.36305\\
          & 3     & 314   & 810   & - & 151 & (0.08384) & 314   & 810   & -     & 153 & (0.14706) \\
          & 4     & 285   & 283   & 316 & - & & 283   & 282   & 314   & - & \\
		\midrule
		\multirow{4}[0]{*}{30} & 1     & -     & 727   & 503 & 220 &  & -     & 730   & 502   & 230 &  \\
          & 2     & 36    & -     & 72 & 10 & 6.69000 & 33    & -     & 65    & 10 & 9.03224\\
          & 3     & 315   & 809   & - & 152 & (0.08246) & 315   & 816   & -     & 160 & (0.02886) \\
          & 4     & 285   & 283   & 315 & - & & 275   & 283   & 307   & - & \\
		\bottomrule
    \end{tabular}}%
  \label{tab:goffituj4}%
\end{table}%

\end{document}